# Conductance of partially disordered graphene: Crossover from temperature-dependent to field-dependent variable-range hopping

C Y Cheah[1], C Gómez-Navarro[2], L C Jaurigue[1] and A B Kaiser[1,3]

[1]MacDiarmid Institute for Advanced Materials and Nanotechnology, School of Chemical and Physical Sciences, Victoria University of Wellington, PO Box 600, Wellington 6140, New Zealand
[2]Dpto de Fisica de la Materia Condensada, Universidad Autonoma de Madrid, 28049, Madrid, Spain

[3]E-mail: Alan.Kaiser@vuw.ac.nz

**Abstract.** We report and analyze low-temperature measurements of the conductance of partially disordered reduced graphene oxide, finding that the data follow a simple crossover scenario. At room temperature, conductance is dominated by two-dimensional (2D) electric field-assisted, thermally-driven (Pollak-Riess) variable-range hopping (VRH) through highly-disordered regions. However, at lower temperatures $T$, we find a smooth crossover to follow the $exp(-E_0/E)^{1/3}$ field-driven (Shklovskii) 2D VRH conductance behaviour when the electric field $E$ exceeds a specific crossover value $E_C(T)_{2D}=(E_a E_0^{1/3}/3)^{3/4}$ determined by the scale factors $E_0$ and $E_a$ for the high-field and intermediate field regimes respectively. Our crossover scenario also accounts well for experimental data reported by other authors for three-dimensional disordered carbon networks, suggesting wide applicability.

## 1. Introduction

Graphene is now finding a multitude of applications for many different purposes [1]. In applications involving electronic conduction, the presence and extent of disorder often play an important role in determining the conduction mechanism. Chemically-derived graphene [2,3], involving the formation of graphene oxide followed by chemical reduction to graphene, is a promising route that has found various electronic applications [4]. This method typically leads to the presence of disordered regions [5] which can, for example, be fashioned to produce active regions as well as transport barriers in the same sheet of graphene [6].

      Measurements of the temperature-dependent conductance $G(T)$ of disordered reduced graphene oxide (rGO) made by Gómez-Navarro *et al* [5] for small applied electric fields followed the Mott relation [7] for variable-range hopping (VRH) in the low-field limit,

$$G(T) = G_1 \exp\left(\frac{-T_0}{T}\right)^p, \qquad (1)$$

with best fit for $p=1/3$ as expected for two-dimensional (2D) VRH. In general, $p=(d+1)^{-1}$ where $d$ is the spatial dimensionality of the sample. The temperature dependence of the pre-factor $G_1$ is often neglected [8] in comparison with the temperature dependence of the exponential term. The Mott parameter $T_0$ for 2D conduction [9],

$$T_0 = \frac{13.8}{k_B N(\varepsilon_F) L^2}, \qquad (2)$$

depends [7] on the density of states $N(\varepsilon_F)$ near the Fermi level and their localization length $L$. $k_B$ is the Boltzmann constant.

Gómez-Navarro *et al* [5] also made measurements on rGO as the applied electric field $E$ was increased and found that the conductance showed a significant increase above the zero-field limit given in equation (1). The conductance near room temperature [10] increased according to the exponential of the applied field $E$, as predicted by the Pollak-Riess expression [11] for field-assisted, thermally-activated VRH in 2D:

$$G(T,E) = G_1 \exp\left(\frac{-T_0}{T}\right)^p \exp\left(\frac{E}{E_a}\right), \qquad (3a)$$

where

$$E_a = \frac{k_B T}{0.18 r(T) e}, \qquad (3b)$$

$r(T)$ is the mean hopping length at temperature $T$ and $e$ is the electronic charge.

In the present paper, we report and analyze additional unpublished data at a temperature of 2 K for voltages that enable us the follow the full crossover from thermally-activated VRH at low electric fields to purely field-driven conduction at high fields. We show that the experimental data are well described by a smooth transition between the Pollak-Riess intermediate-field expression (3) and the high-field expression of Shklovskii [12]:

$$G(E) = G_2 \exp\left(\frac{-E_0}{E}\right)^p, \qquad (4)$$

where conduction is determined by the electric field rather than by temperature. The pre-factor $G_2$ is the conductance at very high fields. The field scale parameter $E_0$ is predicted to be related to the temperature scale parameter $T_0$ in equation (2) by the relation:

$$eE_0 r(T) \approx c k_B T_0, \qquad (5)$$

with the coefficient $c$ predicted to be of the order of unity [12], i.e. the energy difference over a mean hopping distance $r(T)$ of an electron, in an electric field equal to the field scale parameter $E_0$, is expected to have a similar magnitude to the analogous thermal energy $k_B T_0$, where $T_0$ is the temperature scale parameter in the low field limit (equation (2)).

VRH is a common transport mechanism in disordered materials (see e.g. Refs. [13-16]) that have a substantial density of localized states close to the Fermi level, with spatially overlapping wave functions. For materials in which the density of states at the Fermi level is suppressed by Coulombic electron-electron interactions, VRH is expected to follow the Efros-Shklovskii [17,18] model – i.e. with conductance proportional to $exp(T^{-1/2})$ at low applied fields $E$ (for both 2D and 3D materials), and with conductance proportional to $exp(E^{-1/2})$ at high fields – rather than the standard Mott model [7]. The former behaviour has been observed in 3D CdSe thin films [19] and very recently for disordered graphene [20]. We note that our rGO samples differ from those of Lo *et al* [20] in that our samples are only partially disordered with highly conducting crystalline regions making up a large fraction of the sample. While Lo *et al* [20] found in their highly disordered samples that VRH followed Efros-Shklovskii behaviour, i.e. the factor $p=1/2$, as also found by Yu *et al* [19], our partially disordered samples follow [5,10] the standard non-interacting 2D VRH behaviour where $p=1/3$.

A slightly different conduction mechanism is reported in 3D GeAu thin films wherein a crossover occurs from $exp(T^{-1/2})$ to $exp(E)$ as the field is increased [21] (i.e. the field-dependence appears to follow the Pollak-Riess intermediate-field behaviour rather than reaching the high-field limit). In 3D carbon networks, a crossover from $exp(T^{-1/4})$ to $exp(E^n)$ behaviour, wherein $n$ scales from zero to one as the field is increased, has been reported [22]. In the final section of this paper, we shall present our interpretation of the experimental results of reference [22], which can be very well described by a crossover between the Pollak-Riess and Shklovskii expressions (3) and (4) respectively.

## 2. Experimental details

To obtain rGO monolayers, graphite flakes were first oxidized via the Hummers method [2]. The resulting graphite oxide were dispersed in water and subsequently deposited on $SiO_2$(300nm)/Si substrates. These surfaces had been previously provided with alignment marks and treated with aminopropyltrietoxysilane. Deposition time was adjusted in order to obtain a ~ 25% coverage. Once deposited, the GO flakes were chemically reduced under hydrogen plasma with conditions described in reference [5], at which conductivity is maximized. Single layers of 1 nm topographic height were located by Atomic Force Microscopy and contacted with Ti/AuPd electrodes by e-beam lithography. Scanning tunnelling microscopy (STM) analysis showed that the disordered regions in the graphene samples were of mean size ~ 6 nm [5]. Typical monolayer lateral size of the samples was 2 μm$^2$ and the channel length was 500 nm. Electrical transport measurements in a field effect configuration were performed in an Oxford Instrument cryostat under a low pressure of Helium.

## 3. Results and discussion

*3.1 Conduction behaviour of partially disordered graphene*
Figure 1 shows the field dependence of rGO conductance with additional low temperature data at 2 K that we have not previously published. The data in figure 1 are for zero gate voltage ($V_g$=0), i.e near the charge neutrality point. Meanwhile, figure 2 shows data for conduction by electrons (corresponding to the positive applied gate voltage $V_g$) and hole conduction (negative $V_g$) which are qualitatively similar to the $V_g$=0 data but with substantially larger magnitudes of conductance due to the much larger carrier densities. As noted in our earlier paper [10], there is a considerably greater conductance when holes are the majority charge carriers compared to electrons.

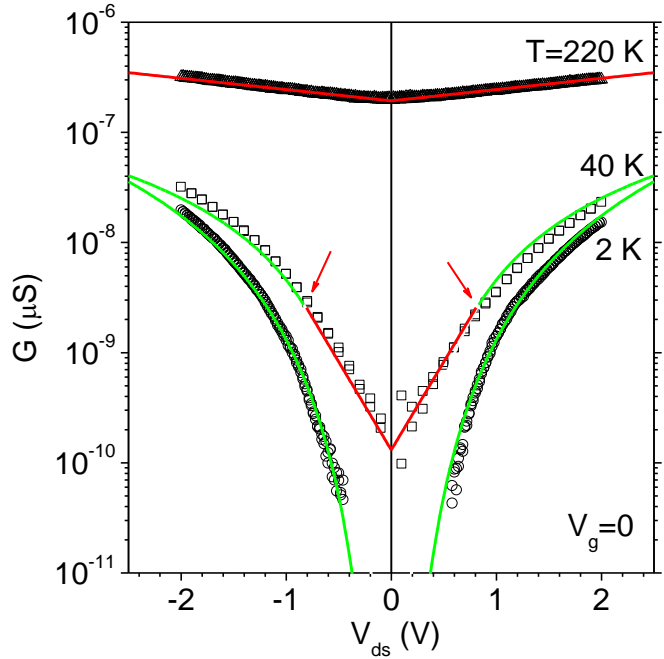

**Figure 1.** Semi-logarithmic plot of the measured conductance (symbols) of disordered graphene versus drain-source voltage $V_{ds}$ at three temperatures *T* as indicated (at zero applied gate voltage). Fits to intermediate-field 2D VRH, equation (3), are shown in red (for *T*=220 K and at low fields below the arrows for *T*=40 K) while fits to high-field 2D VRH, equation (4), are shown in green (for *T*=2 K and at high fields for *T*=40 K). The crossover between the two regimes at temperature 40 K, given by equation (6), is indicated by red arrows.

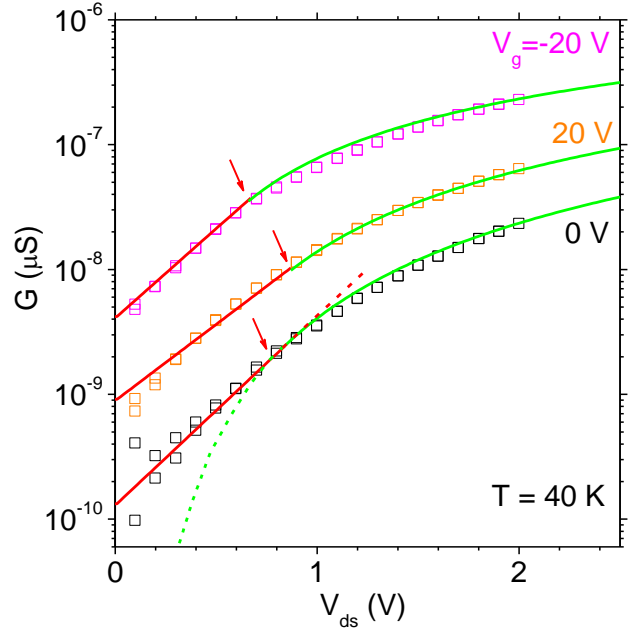

**Figure 2.** Semi-logarithmic plot as in figure 1 showing the different magnitudes of conductance at $T$=40 K (for positive bias voltages only) for different gate voltages corresponding to hole conduction ($V_g$=-20 V), electron conduction ($V_g$=+20 V) as well as near the charge neutrality point ($V_g$=0). The dashed lines for zero gate voltage are representative extrapolations of the fits beyond the appropriate regimes of intermediate-field and high-field VRH, demonstrating the crossover behaviour.

In figure 1, the field-assisted, temperature-activated VRH appear as straight lines in this log-linear plot, with the magnitude of the slope reflecting the importance of the field-assistance term (i.e. the $exp(E/E_a)$ factor) in the Pollak-Riess expression (3). At $T$=220 K, there is a small positive slope corresponding to weak field assistance, but at 40 K, there is a much larger exponential field assistance factor at low bias voltages. The upper limits of the linear Pollak-Riess behaviours are indicated by red arrows in the figure.

At $T$=2 K, we find no evidence of an exponential increase with the field, i.e. there is no linear section of the data at this temperature. Instead, we find that the data are in very good agreement with Shklovskii's high-field VRH expression (4) as shown by the fit in the figure. To check this further, we plot in figure 3 the conductance versus $V^{1/3}$, where $V$ is the bias voltage, finding a very good agreement over a wide range of fields except for small deviations at the lower fields where the high-field expression is no longer applicable. This power law verifies that the hopping is 2D. Fits to the 3D expression $exp(V^{1/4})$ as well as the Efros-Shklovskii expression with $p$=1/2 are significantly poorer. It is clear that at this low temperature, purely field-driven VRH is the dominant mechanism. In contrast, there is no sign of this purely field-driven VRH in conductance at T=220 K. From figures 1 and 2, it can be seen too that the high-field VRH expression (4) gives a good fit to the data at 40 K at higher fields, down to the field values indicated by the red arrows.

If the electric field were uniform throughout the sample, one could calculate the mean hopping length $r(T)$ at temperature 40 K from the Pollak-Riess VRH fits to equation (3). The average electric field can be obtained from the data in figure 1 using the distance between electrodes of 1.1 µm. We obtain mean hopping lengths of order 60-80 nm for the different gate voltages. These hopping lengths are considerably larger than the disordered regions of mean size ~ 6 nm inferred from STM data on the graphene samples [5]. This illustrates that the electric field is instead highly non-uniform and largely concentrated in the disordered barriers [10].

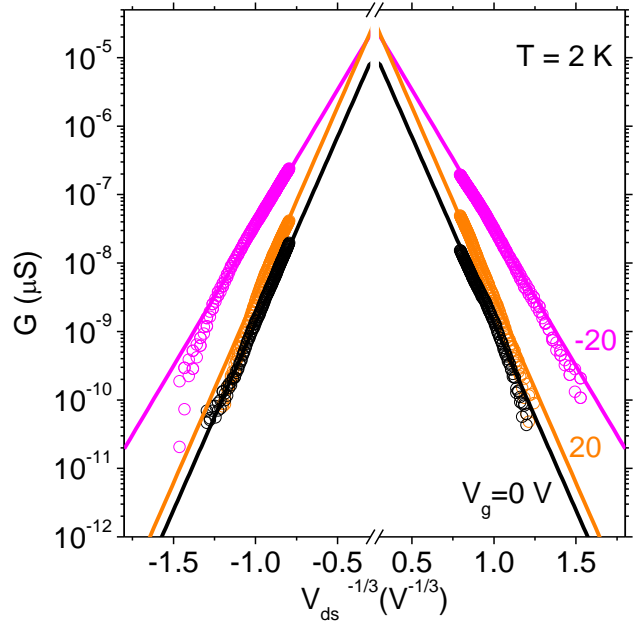

**Figure 3.** Semilogarithmic plot of conductance $G$ versus the inverse one third power of drain-source voltage $V_{ds}^{-1/3}$ at temperature 2 K, for zero and non-zero gate voltages in disordered graphene. Experimental data are shown as symbols, while the lines show fits to the high-field VRH conductance expression, equation (4).

Figure 2 presented conductance data of the rGO at temperature 40 K for zero and non-zero gate voltages. At this intermediate temperature, a crossover between intermediate-field and high-field 2D VRH is evident as indicated by the arrows. The dashed lines for the zero gate voltage data are representative extrapolations beyond the appropriate regimes of intermediate-field and high-field VRH, which demonstrate the clear change in behaviour and so the need for our crossover scenario.

The simplest scenario for the transition between these regimes, starting from the high-field limit, is as follows: as the applied field $V_{ds}$ decreases for a given temperature $T$, the extrapolation of the field-driven Shklovskii term, equation (4), decreases at an increasing rate on the log-linear plot of $G(V_{ds})$. The thermally-activated Pollak-Riess conductance term, equation (3), also decreases albeit at a slower rate. Hence the Pollak-Riess term dominates until in the limit of very low electric field, where the conductance has decreased to the value given by Mott's expression (1) for thermally driven VRH (i.e. the intercept on the axis in figure 1), with the field-activated conduction negligible.

In the Pollak-Riess regime, log $G$ increases linearly as $V_{ds}$ increases. If the field dependence of $G$ retains the Shklovskii high-field form as $V_{ds}$ decreases from the highest fields, there would be a smooth crossover to the Pollak-Riess expression where the slopes of the two expressions are the same. It is easy to show that the equal slopes of the Shklovskii high-field and Pollak-Riess intermediate-field expression for a given temperature $T$ occur at electric fields $E_C$, given by:

$$E_C(T) = \left(pE_a E_0^p\right)^{(p+1)^{-1}}. \qquad (6)$$

We note that only the two experimental fitting parameters, $E_a$ and $E_0$, are needed to calculate the crossover field $E_C$. For our 2D case where $p=1/3$, $E_C(T)_{2D}=(E_a E_0^{1/3}/3)^{3/4}$. For rGO at temperature of 40 K and $V_g=0$, we obtain using equation (6) a crossover voltage of 0.84 V, which matches the crossover point arrowed in figure 1.

*3.2 Application to experimental conductivity data of 3D carbon networks*

We now illustrate how our scenario also applies for the case of 3D VRH in disordered carbon. We plot in figure 4 fits to the conductivity of self-assembled carbon networks made by Govor *et al* [22]. These networks consist of disordered carbon connections with width and height about 100 nm, arranged in a planar pattern of micron-sized irregular hexagons and pentagons.

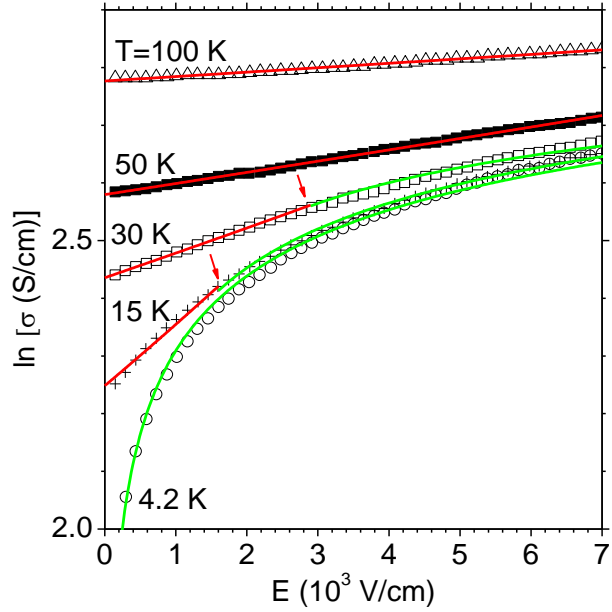

**Figure 4.** Plot of natural logarithm of conductivity versus applied field *E*. Fits of our crossover scenario (intermediate-field 3D VRH, equation (3), in red with the high-field 3D VRH, equation (4), in green) to the experimental data of Govor *et al* [12] (black symbols) show a remarkable similarity to the pattern for disordered graphene. Crossovers between the two regimes at two intermediate temperatures are indicated with red arrows.

In fact, plotting the field-dependent conductance measured by Govor *et al* with similar log-linear axes to those we used for the data of disordered graphene in figures 1 and 2, we notice a striking similarity: at higher temperatures (50 and 100 K) there is an exponential increase of conductance with the electric field *E*, while at very low temperature (4.2 K) the high-field expression of Shklovskii gives an excellent description of the behaviour. At intermediate temperatures (15 and 30 K), there is a smooth crossover from Pollak-Riess to Shklovskii VRH (marked by arrows) that is remarkably similar to the behaviour in disordered graphene, suggesting that this combination of the Mott, Pollak-Riess and Shklovskii expressions gives a very good description of VRH driven by both thermal and electrical energy. Using equation (6) from our model, we calculate the crossover field at temperature 30 K to be $2.9 \cdot 10^3$ V/cm, and at temperature 15 K to be $1.6 \cdot 10^3$ V/cm.

Our analysis shows that the data of Govor *et al* can be interpreted solely within the framework of temperature- and electric field-dependent 3D VRH. In contrast, Govor and colleagues had suggested a model involving a conductivity crossover from the 3D Mott VRH behaviour *exp($T^{-1/4}$)* to *exp($E^n$)* where *n* increased from 0 to 1 as electric field and temperature increased.

*3.3 Relation between thermally activated and purely field driven VRH*

We now use our analyses of both temperature-dependent and field-dependent VRH to evaluate the dimensionless parameter $c$ in Shklovskii's prediction, equation (5), for the relation between the temperature scale factor $T_0$ in the Mott relation (equation (2)), the field scale factor $E_0$ in Shklovskii's VRH theory (equation (4)), and the field-assistance parameter $E_a$ in the Pollak-Riess expression (equation (3)). All three parameters can be evaluated for the rGO data at 40 K in figure 2 where there is a crossover from Pollak-Riess to Shklovskii VRH behaviour as the field increases. Likewise, we can evaluate parameter $c$ for the 3D carbon networks of Govor *et al* [22] at temperatures 30 K and 15 K (figure 4). Our results are summarized in table 1.

**Table 1.** Fit parameters for the Mott, Shklovskii and Pollak-Riess equations and the derived values of the dimensionless ratio $c$ for 2D rGO (all at $T$=40 K) and 3D carbon networks.

| Material | Data set | $T_0$ (K) | $E_0$ (V/m) | $E_a$ ($10^5$ V/m) | $c$ |
|---|---|---|---|---|---|
| 2D rGO | $V_g$=-20 V | 1.7 | 176 | 2.80 | 0.08 |
|  | $V_g$=+20 V | 1.9 | 373 | 3.03 | 0.1 |
|  | $V_g$=0 | 1.6 | 647 | 2.42 | 0.4 |
| 3D C net | $T$=30 K | 0.026 | 506 | 57.69 | 0.6 |
|  | $T$=15 K | 0.045 | 1303 | 21.74 | 1.1 |

It must be noted that the values of fit parameters were obtained assuming the electric field to be uniform over the sample. The rGO results for $c$ are consistent with Shklovskii's estimate, i.e. that it is of order unity, if account is taken of the non-uniform electric field. As mentioned above, it was estimated [10] that for the case of heterogeneous disorder in rGO, the electric field in the barrier regions where hopping occurs is about an order of magnitude larger than the mean field since there is little potential drop across the highly conducting regions that make up most of the conduction path. Hence it follows that the derived values of $c$ are very approximately an order of magnitude smaller than the actual values for hopping in the disordered regions that contributes most of the resistance.

In addition, this field non-uniformity is less significant for the data set near the charge neutrality point where the carrier density is low and the conducting regions have lower conductivity than in the other cases, and thus we propose that for rGO the result $c$=0.4 is the best representative estimate of the true value of the parameter $c$.

The more homogeneously disordered 3D networks of Govor *et al*, on the other hand, should yield more accurate values for $c$. The fact that these values are close to 1 is strong support for Shklovskii's theoretical prediction that $c$ is of order unity.

## 4. Conclusion

In this paper, we have presented experimental data for conduction in partially disordered graphene that show a full range of variable-range hopping behaviour from temperature-activated hopping (with field assistance as the applied voltage increases) to purely field-driven hopping. We have emphasized the transition between these two regimes. Our conclusions are:

(a) The experimental data follow Pollak-Riess behaviour up to intermediate fields (equation (3)), and Shklovskii behaviour (equation (4)) at high fields. We found no evidence for an effect of Coulomb interactions in our data, which instead followed the standard behaviour for 2D hopping conduction.

(b) We find that the transition between field-assisted, temperature-activated hopping to purely field-driven hopping can be well described by a simple crossover scenario when the slope of the log $G$ versus $E$ fits on either side of the crossover field $Ec$ have the same slope (figure 2). This matching of slope corresponds to the smooth transitions observed, and yields the expression of equation (6), $E_C(T)_{2D}=(E_a E_0^{1/3}/3)^{3/4}$, for the crossover field $E_c$ in terms of the field scale factors $E_a$ for field-assisted, temperature-activated hopping and $E_0$ for purely field-driven hopping. Our results highlight the need for further theoretical studies of the transition regime where the contributions of thermal activation and field-activated hopping are both significant.

(c) The values of the fitting parameters are consistent with the predicted relation (equation (5)) between the thermal-activation scale factor $T_0$ and the analogous scale factor $E_0$ for the purely field-driven case, with increasingly good agreement as the hopping becomes more uniform over the sample.

(d) We have also shown that our crossover scenario provides a natural explanation for the data of Govor *et al* [22] for 3D disordered carbon networks, indicating the wide applicability of our scenario.

**Acknowledgments**

We thank Ravi Sundaram and Marko Burghard for their collaboration on the experimental project. CYC would like to thank the MacDiarmid Institute for Advanced Materials and Nanotechnology for financial support.